# How to create an interface between UrQMD and Geant4 Toolkit

Khaled Abdel-Waged[1,2],
Nuha Felemban[1], V.V. Uzhinskii[3]

[1]Physics Department, Faculty of Science, Umm Al-Qura University, Saudi Arabia
[2]Physics Department, Faculty of Science, BenhaUniversity, Egypt.
E.mail:khelwagd@yahoo.com.
[3]The European organization for nuclear research (CERN)

## Abstract

An interface between the UrQMD-1.3cr model (version 1.3 for cosmic air showers) and the Geant4 transport toolkit has been developed. Compared to the current Geant4 (hybrid) hadronic models, this provides the ability to simulate at the microscopic level hadron, nucleus, and anti-nucleus interactions with matter from 0 to $10^{12}$ eV/A with a single transport code. This document provides installation requirements and instructions, as well as class and member function descriptions of the software.

# Contents

# 1 Introduction

Geant4 [1] use many transport codes to generate non-elastic reactions in macroscopic targets. The motivation was to provide a satisfactory level of description of final state hadron production for accurate and comprehensive simulations of particle detectors used in modern particle and nuclear physics experiments.

In Geant4 [1] there are mainly 4 codes which are frequently applied for simulating N-body collisions within the hadronic cascade models, namely, Binary [2], Bertini [3], FTF[4] and QGS[5]. The Geant4 Binary and Bertini cascade models implement large resonance collision term but multiparticle production through string formation and fragmentation is not considered, which sets the upper limit of model applicability in nucleon induced reactions to about 5 GeV/c incident momentum. In contrast to **Binary** and **Bertini** cascade models, **FTF** and **QGS**, takes multiparticle production into account, to handle interaction for particles above about 15-20 GeV/c.

For a reliable description of secondaries produced in high and low energy domains, Geant4 invoke **Binary (Bertini)** in the **FTF(QGS)** models. The corresponding hybrid model is called FTFB(QGSB).

To improve the simulation, it is worthwhile to supplement Geant4 toolkit with a single transport code which takes into account both resonance and multiparticle production. The UrQMD (Ultra relativistic Quantum Molecular Dynamics) model [6], which is first released in 1998, is designed to simulate (ultra) relativistic heavy ion collisions in a wide energy range from Bevalac and SIS up to AGS, SPS and RHIC. Thanks to the UrQMD collaborators, the UrQMD is provided as open source Fortran code on its website.

**Geant4-UrQMD** interface offers interactions of Protons, neutrons, pions and kaons (and their antiparticles) as well as light and heavy ions interactions with matter. It gives type, mass, energy, and momentum of all secondary particles that are created in matter. Similar to CORSIKA-UrQMD [7] interface, the upper energy limit is recommended to be about several 100 GeV. For the simulation, the appropriate **UrQMD** cross sections (for determining the mean free path between interacting particles) are used. However, the probability for an interaction with matter at a given point is computed by the Geant4 cross sections classes.

This document provides installation requirements and instructions of the Geant4 -UrQMD interface. More details about validation with Geant4 cascade models can be found in Ref.[8]. This interface is sponsored by King Abdulaziz City for Science and Technology /National Center for Mathematics and Physics- Saudi Arabia under the contract number **31-465**.

## 2 Brief description of UrQMD model

The UrQMD model version 1.3cr (for cosmic air shower) is a Monte Carlo event generator for simulating hadron-hadron, hadron-nucleus, nucleus-nucleus interactions from 0 to $10^6$ MeV. The main goal is to gain understanding about Physical phenomena within a single transport mode. As such, it is relevant both to high-energy hadron/ion accelerators and cosmic radiation environments.

The UrQMD model is based analogous principles as the quantum molecular dynamic (QMD) model and the relativistic QMD (RQMD): the mean field potential applied to hadrons is treated similar to QMD, while the treatment of the collision term is similar to RQMD.

Compared with Geant4 simulation codes (e.g., Binary, Bertini, FTF, QGSM…), UrQMD has the following improvements to its physics (see Ref.[6] for full description):

- The collision term of the UrQMD treats 55 different baryon species (including nucleon, $\Delta$, $\Lambda$, $\Sigma$, $\Xi$ and $\Omega$ and their resonances with masses up to 2.25 GeV/$c^2$ ) and 32 different meson species (including strange meson resonances with masses up to 1.9 GeV/$c^2$) as well as their antiparticles and explicit isospin projected states. For excitations with higher masses than 2 GeV/$c^2$ the string model is used (The exited hadronic states can either be produced in inelastic *hh* collision, resonance decays or string decays).
- All of these hadronic states can propagate and re-interact in phase-space.
- A mean field potential can be applied for the scattered nucleons.

- An analytic expression for the differential cross section of in-medium NN elastic scattering derived from the collision term of the relativistic Boltzmann-Uehling-Uhlenbeck (RBUU) equation is used to determine the scattering angles between the outgoing particles in hh collisions.
- An introduction of the formation time and leading pre-hadron effects. For the leading pre-hadrons, a reduced cross section is adopted according to the constituent quark model.

UrQMD-1.3cr is freely available from the website http://urqmd.org/.

# 3 Installation requirements

The Geant4 interface to UrQMD-1.3cr (for cosmic air showers) uses the following software tools and packages:

1. UrQMD-1.3cr (available at http://urqmd.org/ website)
2. The following Geant4 versions:
   - *ver.9.3.p01* or *ver.9.5*

The interface and original UrQMD-1.3cr source code have been compiled and tested using:

- *gcc 4.1.2 with g77 (FORTRAN77), gfortran (FORTRAN95) and GNUmake)*

Operating systems used for this test is *: Red Hat Linux 4.1.2-64*

# 4 Interface design

The use of UrQMD-1.3cr physics in Geant4 has resulted in the introduction of a new event model (**G4UrQMD1_3Model**).

## 4.1 Logical description of G4UrQMD1_3Model

The general steps in the logical sequence of the design are:

### 1. Instantiation of G4UrQMD1_3Model

Upon instantiation of the UrQMD-1.3cr interface, the class constructor initialises variables associated with the class and the UrQMD-1.3cr FORTRAN model. The latter involves initialisation of a series of common-block variables from the C++ code, mimicking the initialisation process performed by the FORTRAN code when used in standalone mode.

### 2. Generation of the interaction events

In order to generate interaction events, the **ApplyYourself** member function of Geant4 is executed. The details of the projectile and target are determined (e.g., mass number, charge, projectile energy…).

If no data exist for the specific projectile-target combination, a “waning” is declared, and the program is stopped.

## 4.2 Class structure for model

The interface comprises **G4UrQMD1_3Model** class, which is derived from **G4HadronicInteraction**, and is defined within the Geant4 user physics list if access to **UrQMD1_3** physics is required. It controls initialisation of **UrQMD-1.3cr** through common block variables.

A small FORTRAN subroutine is included with the G4UrQMD distribution. This is required to ensure that block data initialisation in G4UrQMD-1.3c is performed (through FORTRAN EXTERNAL statements). Specifically the new FORTRAN subroutine is:

- **g4urqmdblockdata.f**

# 5 Instructions for software installation in geant4-9.3

## 5.1 Installing UrQMD-1.3cr and building the libG4hadronic_urqmd13 library

Copy the file **g4urqmd-1.3cr.2012.tar** ( available at https://twiki.cern.ch/twiki/bin/view/Sandbox/Geant4UrQMDinterface )

to the directory:

**cd $G4INSTALL/source/processes/hadronic/models**

Unzip or untar the files, *e.g.*:

**tar –xvzf g4urqmd-1.3cr.2012.tar**

This creates the directory **UrQMD13**, which looks like:

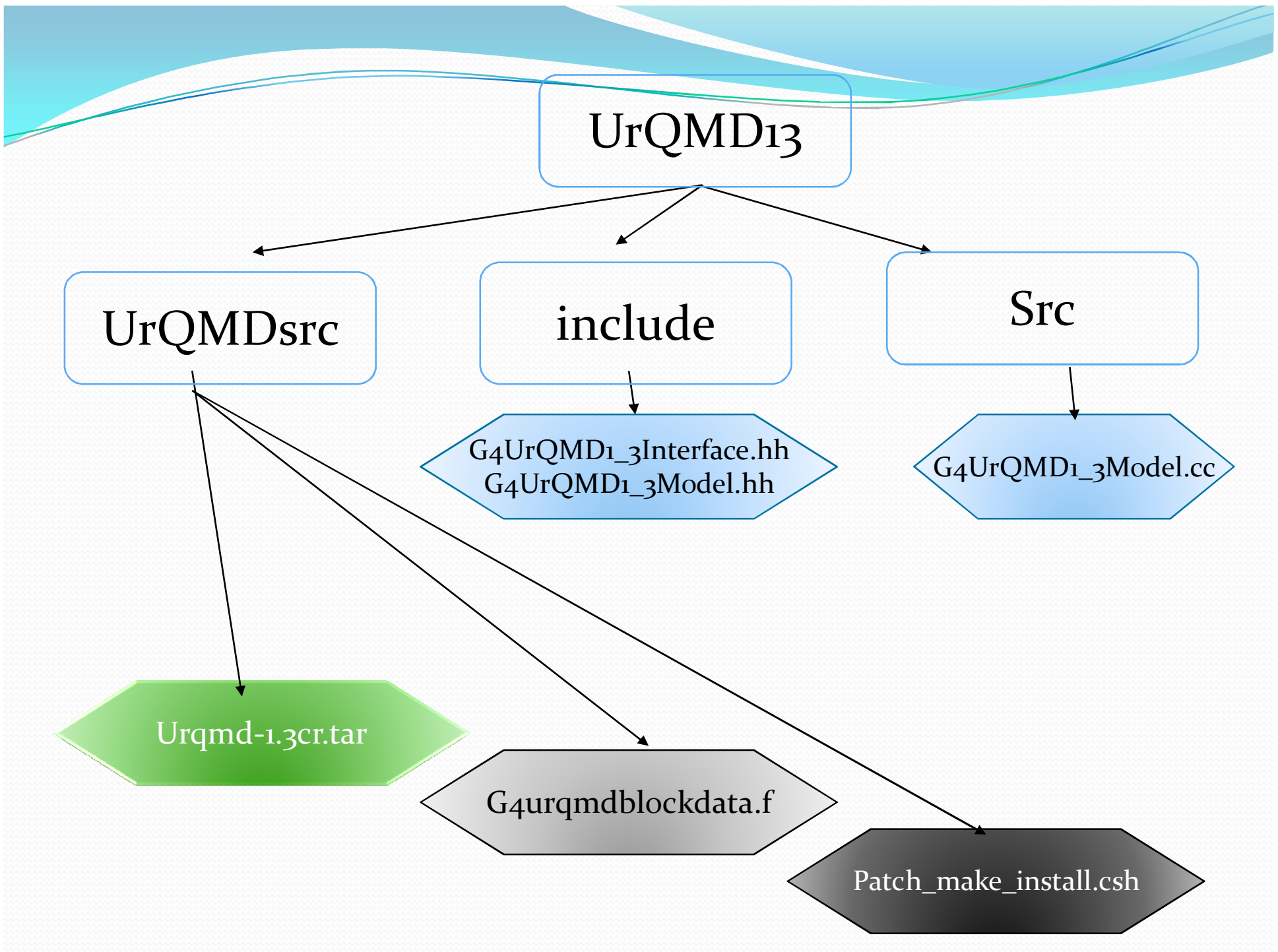

Note that the header and C++ files (**G4UrQMD1_3Interface.hh & G4UrQMD1_3Model.hh)** and **G4UrQMD1_3Model.cc** are in **UrQMD13/include** and **UrQMD13/src** respectively.

Change directory:

**cd UrQMD13/UrQMD13src**

If the file **urqmd-1.3cr.tar** doesn't already exist in the directory, copy it from the web-site: http://urqmd.org/ .

Execute the script **patch_make_and_install.csh** from a **csh/tcsh** environment. This initiates 2 steps:

1- All the relevant FORTRAN files are compiled in **UrQMD13src**, using

**FC= g77**

**LD=g77**

**FFLAGS=-g –Wall –Wsurprising –fbound –check**

**LDFAGS= -g**

2- The object files are copied to the directory:

**$G4TMP/$G4SYSTEM/G4hadronic_UrQMD13**

Change directory and execute **make** on the C++ source:

**cd ..**

**make**

This will build the library:

**$G4LIB/$G4SYSTEM/libG4hadronic_UrQMD13.a**

but note that the **Gnumake** file of your application must explicitly identify this library and the **UrQMD1_3** headers. So in **GNUmake** file for you application, please add the following line:

**CPPFLAGS += -I$(G4BASE)/processes/hadronic/models/UrQMD13/include**

Note that if you perform:

**make clean**

from **UrQMD13** directory, this **deletes** the FORTRAN-compiled object files which were from **UrQMD13/UrQMD13src** to **$G4TMP/$G4SYSTEM/ libG4hadronic_urqmd13**. Therefore, you should either re-copy the files, or re-run the script

**patch_make_and_install.csh** before performing a make in directory **UrQMD13**.

## 5.2 The main () program

Geant4 does not provide the **main().** In the **main()**, one has to construct **G4ProcessManager**, in order to process a run for the interface.

Ensure that the **GNUmake** file for you application (in "**main**" directory) includes the following lines:

**CPPFLAGS += -I$(G4BASE)/processes/hadronic/models/UrQMD13/include**

**EXTRALIBS += -L$(G4INSTALL)/lib/$(G4SYSTEM) –lG4hadronic_UrQMD13**

*If you want to compile the Geant4-UrQMD interface from your "main" directory, please add these lines in* **GNUmake file***:*

**EXTRA_LINK_DEPENDENCIES:= $(G4INSTALL)/lib/$(G4SYSTEM)/ libG4hadronic_UrQMD13.so**

**$G4INSTALL)/lib/$(G4SYSTEM)/ libG4hadronic_UrQMD13.so:**
**cd $(G4BASE)/processes/hadronic/models/UrQMD13; $(MAKE)**

# 6 Class descriptions (summary)

## Class G4UrQMD1_3Model

### Public member functions

| |
|---|
| `G4UrQMD1_3Model(const G4String& name = "UrQMD1_3");` |
| `virtual ~G4UrQMD1_3Model ();` |
| `virtual G4HadFinalState *ApplyYourself (const G4HadProjectile&,4Nucleus&);` |

### Private member functions

| |
|---|
| `G4int operator==(G4UrQMD1_3Model& right){return (this == &right);}` |
| `G4int operator!=(G4UrQMD1_3Model& right){return (this != &right);}` |
| `void WelcomeMessage () const;` |
| `void InitialiseDataTables ();` |
| `G4bool offshell` |

# 7 Class descriptions (full descriptions)

The following sub-sections describe in detail each of the new or modified classes introduced into Geant4 to treat the interface with UrQMD-1.3cr. The descriptions include the overall purpose of the class, the constructor, destructor, and private/public member functions of the classes, including identification of the arguments and quantities returned. It should be noted that, the ***units of all inputted and returned quantities are default units used in Geant4***, *e.g.* lengths (such as nuclear radii) are stated in millimetres, areas (*e.g.* interaction cross-sections) in square-millimetres, and energies are stated in MeV.

## 7.1 Class G4UrQMD1_3Model

The **G4UrQMD1_3Model** class is the main class driving the UrQMD-1.3cr FORTRAN code and primarily acts as an event generator to Geant4 to determine the final state of an interaction. It is derived from **G4HadronicInteraction** class.

**Public constructors, operators and member functions**

```
G4UrQMD1_3Model(constG4String&name=UrQMD1_3");
```

public constructor which calls **InitialiseDataTables** (). This initialises all of the variables as required by UrQMD-1.3cr and held in FORTRAN common blocks based on the DEFAULT values, *i.e.* largely based on the UrQMD-1.3cr defaults, see CTParam(X) and CTOption(X) from Tables 5 and 6 of the UrQMD user guide (http://urqmd.org/).

**~G4UrQMD1_3Model ();**

Class destructor.

**virtual G4HadFinalState *ApplyYourself (const G4HadProjectile *&theTrack*, G4Nucleus *&theTarget*);**

Member function to determine the **Final State** of the interaction process. The **Final state** is for the results of (and secondaries from) the projectile, described by **theTrack**, and **the target** nucleus, described by theTarget. This member function drives the UrQMD-1.3cr FORTRAN to enter information about an interaction event, and interprets the results into a product vector as a **G4HadFinalState** object.

## Private constructors, operators and member functions

**const G4UrQMD1_3Model& operator=(G4UrQMD1_3Model &right);**

Operator which is required by Geant4 physics lists.

**void WelcomeMessage ();**

Prints to standard output (stdout) the information about the G4UrQMD1_3Model when the class is instantiated.

**void InitialiseDataTables ();**

Initialises the UrQMD-1.3cr variables during the execution of either of the constructors. This is an initialisation process, primarily setting common block variables directly or through block data calls.

**Gbool offshell;**

Option to be fixed by the user:

**true** if nucleons are put "**off-mass-shell**" (default UrQMD option) and **false** if nucleons are "**on-mass-shell**" (recommended by the authors).

Note that this option strongly affects low energy cascade particles.

## 7.2 G4UrQMD1_3Interface: C++ --FORTRAN interface for common blocks and subroutines

Although not representing a class in itself, **G4UrQMD1_3Interface.hh** contains a series of structure definitions in order to permit access to FORTRAN common block variables. For each common block used by the Geant4 interface, there is an associated C/C++ struct. For example, common block **urqmdparams** becomes the variable **urqmdparams** _ of struct type **ccurqmd13urqmdparams**, and variables **real*8 u_elab** and **integer u_at** in that common block are designated as **urqmdparams_.u_elab** and **urqmdparams_.u_at** respectively in the C/C++ interface.

The other function of **G4UrQMD1_3Interface.hh** is to define cross-references between the **UrQMD1.3cr** FORTRAN **subroutines** and C/C++ **subroutines**.

# 8 Changes to usage of geant4-9.5

In version 9.3.p01 (and older ones) the anti-nucleus characteristics and cross sections are not supplied with the geant4 source. Therefore, one has to install the most recent version of geant4 (9.5) in order to allow anti-nucleus interactions using UrQMD code. For installation, one has to use the so called **CMAKE** build system to compile and install the toolkit. **CMAKE** has been chosen to replace the old **Configure/Make** system. For more information on **CMAKE** itself, please visit http://www.cmake.org/.

## 8.1 Building and installing CMAKE

Unpack the **cmake-2.8.7.tar.gz** to a location of your choice. For illustration only, assume it is unpacked in a directory **/home/path,** so that the CMAKE source sits in the directory

**$ /home/path/make**

**$tar –xvzf cmake-2.8.7.tar.gz**

To configure the build:

**$./bootstrap**

***(The following step is not written in the CMAKE manual)***

**$./configure - - prefix=/home/path/cmake**

**$make**

**$make install**

*Here the Configure variable prefix is used to set the install directory, the directory under which CMAKE libraries, binaries and headers are installed.*

Finally you will get the directory “cmake” which contains:

+-cmake

+-Bin/

|+-ccmake

|+-**cmake**

|+-cpack

|+-ctest

+-cmake-2.8.7/

+-doc/

|+-cmake-2.8/

+-man/

|+-man1/

+-share/

|+-alocal/

|+-cmake-2.8/

To make sure that everything is O.K., please write

**$ cmake -version**

You will get

**$cmake version 2.8.7**

To run **cmake** in any path terminal, you have to add few lines in your **.bash_profile** as follows.

First open your **.bash_profile**:

**$nano .bash_profile**

Write

**PATH=/home/path/cmake/bin:$PATH**

**LD_LIBRARY_PATH=/home/path/cmake/lib:$LD_LIBRARY_PATH**

Save your **.bash_profile.**

Finally, execute:

**$source .bash_profile**

## 8.2 Building and installing geant4.9.5

First make sure that there is no previous version of geant4 in your system. To do this, please type:

**$ printenv | grep G4**

Maybe you find something in your **.bash_profile**. If you do, go to main terminal and write

**$nano .bash_profile**

Erase or mark anything starts with **G4**.

To ensure that everything is O.K., write:

**$source .bash_profile**

**$printenv | grep G4**

If you find no "**G4**" proceed as follows:

Unpack the **geant4.9.5.tar.gz** to the location **/home/path/geant4,** so that the geant4 "unpacked files" sits in the subdirectory **/home/path/geant4/geant4.9.5**

The next step is to create a directory to configure and run the build and store the build products. This directory should be alongside your source directory (geant4-9.5):

**$mkdir geant4.9.4-build**

**$ls**

**+-geant4/ +-geant4.9.5/ +-geant4.9.5-build/**

To configure and build, change into the build directory:

**$cd geant4.9.5-build**

Since geant4 source is in **/home/path/geant4/geant4.9.5,** it is recommended to install it in **/home/path/geant4/geant4.9.5-install**

by typing:

**$cmake -DCMAKE_INSTALL_PREFIX=/home/path/geant4/geant4.9.5-install ../geant4.9.5**

When complete, do

**$make**

**$make install**

Finally, you will get **"geant4.9.5-install"** directory which contains

**+-geant4.9.5-install**

**+-bin/**

|+-geant4.csh

|+-geant4.sh

|+-geant4-config

**+-include/**

**|+-Geant4/**

|+-globals.hh

|+-G4ParticleDefinition.hh

|+-G4AntiDeuteron.hh

|+-…

|+-CLHEP/

|+-tools/

**+-lib/**

|+-libG4global.so

|+-…

**+-share/**

**+-Geant4-9.5.0/**

**+-geant4make/**

**|+-config/**

|+- **geant4make.sh**

|+-geant4make.csh

|+…

Finally, make sure that geant4 binaries and libraries are available on your path. This is done by writing (in your **.bash_profile**):

PATH=$PATH:$HOME/geant4_workdir/bin/Linux-g++/

**Source /home/path/geant4/geant4.9.5-install/share/Geant4-9.5.0/geant4make/geant4make.sh**

## 8.3 Building and installing Geant4-UrQMD interface for version 9.5

With version 9.5, you may have problems linking C++ geant4 code against FORTRAN-UrQMD code. In order to resolve these problems it is recommended to use **gfortran** instead of **g77**.

Thus go to the **GNUmake** file in the directory:

**$cd /home/path/geant4/geant4.9.5/source/processes/hadronic/models/UrQMD13**

and make sure that these lines

**FC= g77**

**LD=g77**

**FFLAGS=-g –Wall –Wsurprising –fbound –check**

**LDFAGS= -g**

are replaced with:

**FC=gfortran**

**LD=gfortran**

**FFLAGS=-Fpic**

**LDFLAGS=”-L/usr/lib64 –lgfortran”**

Also, it would be better to abandon the **patch_make_and_install.csh** environment and write few lines in **GNUmake** file:

```
.PHONY: all
all:  urqmd   lib
ifdef G4INCLUDE
CPPFLAGS += -I $(G4INCLUDE)
endif
UrQMDDIRNAME= UrQMDsrc
urqmd:
(cd ${URQMDDIRNAME} && \
tar  -xzf  urqmd-1.3cr.tar.gz && \
cp GNUmakefile urqmd-1.3cr && cp *.f   urqmd-1.3cr &&\
cd urqmd-1.3cr && ${MAKE} FFLAGS=-Fpic   FC=gfortran  LD=gfortran
LDFLAGS=”-L/usr/lib64 –lgfortran”);
```

**(mv ${URQMDDIRNAME}/urqmd-1.3cr/obj_Linux/*.o ${G4TMPDIR});**

This will:

1- Move the object files to the directory

**$/home/path/geant4_workdir/tmp/Linux-g++/G4hadronic_UrQMD13**

2- Create UrQMD library in the directory:

**$/home/path/geant4/geant4-9.5.0/Linux-g++ libGhadronic_UrQMD13.so**

If the linking is O.K. goto **UrQMD13** directory and type:

**$ make includes**

This will move the include files (**G4UrQMD1_3Interface.hh & G4UrQMD1_3Model.hh)** to **Geant4** directory:

**/home/path/geant4/geant.9.5-install/include/Geant4**

Finally in the **GNUmake** file of your "**main()**" application, make sure that this line:

**EXTRALIBS += -L/{G4lib}/${G4SYSTEM} –lG4hadronic_UrQMD13**

is inserted.

# 9 The output file

```
=========================================================
======        Cascade Test (test30) Start       ========
=========================================================
###### Start new run # 0   for 1 events  #####
### NIST DataBase for Materials is used
Material is selected: G4_Zr
Test30Physics entry
Process is created; gen= UrQMD1_3
 ****************************************************************
 Interface to        G4UrQMD_1.3                      activated
 Version number : 00.00.0B          File date : 25/01/12
 (Interface written by Kh. Abdel-Waged et al. for the KACST/NCMP)

 ****************************************************************
 seed:  1097569630
New generator for material G4_Zr Nelm= 1 Nmat= 1 Target element: Zr
Secondary generator <UrQMD1_3> is initialized
Nucleus with N= 94  Z= 40  A(amu)= 9.131840e+01Mass from G4NucleiProperties(GeV)= 8.745283e+01
The particle:  anti_deuteron
The material:  G4_Zr Z= 40 A= 94  Amax= 95
The step:      1.000000e-05 mm
The position:  (0.000000e+00,0.000000e+00,0.000000e+00) mm
The direction: (0.000000e+00,0.000000e+00,1.000000e+00)
The time:      0.000000e+00 ns
energy = 2.000000e+05 MeV    RMS(MeV)= 0.000000e+00
emax   = 2.500000e+01 MeV
pmax   = 2.018669e+05 MeV
### Log10 scale from -2.000000e+00 to 2.000000e+00 in 80 bins
### Histo books 57 histograms in <UrQMD1_3.root>
Histograms are booked output file <UrQMD1_3.root>
----------------------------------------------------------
### factor  = 2.971747e-01### factora = 7.567491e+03### factorb = 2.377397e+03    cross(mb)=
2.377397e+03
Test rotation= (0.000000e+00,0.000000e+00,1.000000e+00)
### 0-th event start
 creation of table, wait-------
 ############################################################
 ##                                                        ##
 ##     UrQMD 1.3    University of Frankfurt               ##
 ##                  urqmd@th.physik.uni-frankfurt.de      ##
 ##                                                        ##
 ############################################################
 ##                                                        ##
 ##     please cite when using this model:                 ##
 ##     S.A.Bass et al. Prog.Part.Nucl.Phys. 41 (1998) 225 ##
 ##     M.Bleicher et al. J.Phys. G25  (1999) 1859         ##
 ##                                                        ##
 ############################################################
 tabnameurqmd.tab
 Generating table...
 (1/7) ready.
 (2/7) ready.
 (3/7) ready.
 (4/7) ready.
 (5/7) ready.
 (6/7) ready.
 (7/7) ready.
 Writing new table...
 O.K.

 end to create  table
UrQMDModel running-------------
 (info) dsigma: calculating constants for ang. dist.
 (info) dsigma: calculation finished
  User=3.225000e+01s Real=3.319000e+01s Sys=3.000000e-02s
###### Save histograms
###### End of run # 0     ######
Next line #exit

      ###### End of test #####
```

**Fig.2 Sample header of the output of Geant4-UrQMD interface**

# 10 G4UrQMD1_3.cc file

```
// ********************************************************************
// * License and Disclaimer                                           *
// *                                                                  *
// * The  Geant4 software  is  copyright of the Copyright Holders  of *
// * the Geant4 Collaboration.  It is provided  under  the terms  and *
// * conditions of the Geant4 Software License,  included in the file *
// * LICENSE and available at  http://cern.ch/geant4/license .  These *
// * include a list of copyright holders.                             *
// *                                                                  *
// * Neither the authors of this software system, nor their employing *
// * institutes,nor the agencies providing financial support for this *
// * work  make  any representation or  warranty, express or implied, *
// * regarding  this  software system or assume any liability for its *
// * use.  Please see the license in the file  LICENSE  and URL above *
// * for the full disclaimer and the limitation of liability.         *
// *                                                                  *
// * This  code  implementation is the result of  the  scientific and *
// * technical work of the GEANT4 collaboration.                      *
// *                                                                  *
// * Parts of this code which have been  developed by Abdel-Waged    *
// * et al under contract (31-465) to the King Abdul-Aziz City for    *
// * Science and Technology (KACST), the National Centre of          *
// * Mathematics and Physics (NCMP), Saudi Arabia.                    *
// *                                                                  *
// * By using,  copying,  modifying or  distributing the software (or *
// * any work based  on the software)  you  agree  to acknowledge its *
// * use  in  resulting  scientific  publications,  and indicate your *
// * acceptance of all terms of the Geant4 Software license.          *
// ********************************************************************
////////////////////////////////////////////////////////////////////////////
////
//
%%%%%%%%%%%%%%%%%%%%%%%%%%%%%%%%%%%%%%%%%%%%%%%%%%%%%%%%%%%%%%%%%%%%%%%%%%%
%
//
// MODULE:           G4UrQMD1_3Model.hh
//
// Version:           0.B
// Date:            25/01/12
// Authors:         Kh. Abdel-Waged and Nuha Felemban
// Revised by:       V.V. Uzhinskii
//                   SPONSERED BY
// Customer:         KACST/NCMP
// Contract:         31-465
//
//
%%%%%%%%%%%%%%%%%%%%%%%%%%%%%%%%%%%%%%%%%%%%%%%%%%%%%%%%%%%%%%%%%%%%%%%%%%%
%
//
#include "G4UrQMD1_3Model.hh"
#include "G4UrQMD1_3Interface.hh"
//------------------------------
#include "globals.hh"
#include "G4DynamicParticle.hh"
#include "G4IonTable.hh"
#include "G4CollisionOutput.hh"
```

```
#include "G4V3DNucleus.hh"
#include "G4Track.hh"
#include "G4Nucleus.hh"
#include "G4LorentzRotation.hh"

#include "G4ParticleDefinition.hh"
#include "G4ParticleTable.hh"

//AND->
#include "G4Version.hh"
//AND<-
//----------------new_anti
#include "G4AntiHe3.hh"
#include "G4AntiDeuteron.hh"
#include "G4AntiTriton.hh"
#include "G4AntiAlpha.hh"
//---------------------------
#include <fstream>
#include <string>

/////////////////////////////////////////////////////////////////////////////
//////

//
G4UrQMD1_3Model::G4UrQMD1_3Model(const G4String& nam)
  :G4VIntraNuclearTransportModel(nam), verbose(0)
{


  if (verbose > 3) {
    G4cout << " >>> G4UrQMD1_3Model default constructor" << G4endl;
  }



//
// Set the minimum and maximum range for the UrQMD model

//   SetMinEnergy(100.0*MeV);
//   SetMaxEnergy(200.0*GeV);
//----------------------------------------new_off shell

//
//   It would be better to turn  "MASS-SHELL" OFF !!
//   because  NUCLEONS should be "ON -MASS-SHELL".
//   However, we turn it on to get default UrQMD options!

     offshell=true;
//----------------------------


//

//
  WelcomeMessage();
//
  CurrentEvent=0;
//

InitialiseDataTables();
```

```
//
}
///////////////////////////////////////////////////////////////////////////
/////
//
// Destructor
//
G4UrQMD1_3Model::~G4UrQMD1_3Model (){}
///////////////////////////////////////////////////////////////////////////
/////

G4ReactionProductVector* G4UrQMD1_3Model::Propagate(G4KineticTrackVector* ,
                              G4V3DNucleus* ) {
  return 0;
}

///////////////////////////////////////////////////////////////////////////
/////
//
// ApplyYourself
//
// Member function to process an event, and get information about the
products.


  G4HadFinalState *G4UrQMD1_3Model::ApplyYourself (
  const G4HadProjectile &theTrack, G4Nucleus &theTarget)
{
//anti_new
//  ------------------define anti_light_nucleus
const G4ParticleDefinition* anti_deu =
  G4AntiDeuteron::AntiDeuteron();

const G4ParticleDefinition* anti_he3=
  G4AntiHe3::AntiHe3();

const G4ParticleDefinition* anti_tri=
  G4AntiTriton::AntiTriton();

const G4ParticleDefinition* anti_alp=
  G4AntiAlpha::AntiAlpha();
//--------------------------------------------------
//
// The secondaries will be returned in G4HadFinalState &theResult -
// initialise this.  The original track will always be discontinued and
// secondaries followed.
//
  theResult.Clear();
  theResult.SetStatusChange(stopAndKill);

  G4DynamicParticle* cascadeParticle=0;
//
//
// Get relevant information about the projectile and target (A, Z,
energy/nuc,
// momentum, etc).
//


  const G4ParticleDefinition *definitionP = theTrack.GetDefinition();
```

```
  const G4double AP        = definitionP->GetBaryonNumber();
  const G4double ZP        = definitionP->GetPDGCharge();
        G4double AT        = theTarget.GetN();
        G4double ZT        = theTarget.GetZ();
//  -----------------------------------------------
      G4int id=definitionP->GetPDGEncoding();  //get particle encoding
// ------------------------------------------------
  G4int AP1 = G4lrint(AP);
  G4int ZP1 = G4lrint(ZP);
  G4int AT1 = G4lrint(AT);
  G4int ZT1 = G4lrint(ZT);
//    G4cout<<"------ap1--=="<<AP1<<"---zp1---=="<<ZP1<<"---id-=="<<id<<
G4endl;
//
//
***************************************************************************
*
// The following is the parameters necessary to initiate Uinit() and
UrQMD()
// ------------------------------------------------------------------------
----
  urqmdparams_.u_sptar=0;  //!0= normal proj/target, 1=special proj/tar
  urqmdparams_.u_spproj=1;  // projectile is a special particle

//new_anti

  if (AP1>1 ||definitionP==anti_deu ||definitionP==anti_he3
            ||definitionP==anti_tri ||definitionP==anti_alp)
      {

      urqmdparams_.u_ap=AP1;
      urqmdparams_.u_zp=ZP1;

      urqmdparams_.u_spproj=0;
      } else if (id==2212) {            //!proton
        urqmdparams_.u_ap=1;
        urqmdparams_.u_zp=1;


      } else if(id==-2212){            //! anti-proton
        urqmdparams_.u_ap=-1;
        urqmdparams_.u_zp=-1;
      } else if(id==2112){              //! neutron
        urqmdparams_.u_ap=1;
        urqmdparams_.u_zp=-1;

      } else if(id==-2112){            //! anti-neutron
        urqmdparams_.u_ap=-1;
        urqmdparams_.u_zp=1;

      } else if(id==211) {              //! pi+
        urqmdparams_.u_ap=101;
        urqmdparams_.u_zp=2;
      } else if(id==-211) {            //! pi-
        urqmdparams_.u_ap=101;
        urqmdparams_.u_zp=-2;
    } else if(id==321) {              //! K+
        urqmdparams_.u_ap=106;
        urqmdparams_.u_zp=1;
    } else if(id==-321) {              //! K-
        urqmdparams_.u_ap=-106;
```

```
        urqmdparams_.u_zp=-1;
    } else if(id==130 || id==310) {     //  ! K0
        urqmdparams_.u_ap=106;
        urqmdparams_.u_zp=-1;
    } else if(id==-130 || id==-310){  // ! K0bar
        urqmdparams_.u_ap=-106;
        urqmdparams_.u_zp=1;
    }  else {

      G4cout << " Sorry, No definition for particle for UrQMD::"<<id<<
"found" << G4endl;

      //AND->
#if G4VERSION_NUMBER>=950
      //New signature (9.5) for G4Exception
      //Using G4HadronicException
      throw G4HadronicException(__FILE__,__LINE__,"Sorry, no definition for
particle for UrQMD");
#else
    G4Exception(" ");
#endif
    //AND<-
      }  //end if id
//---------------------------------------------------------

  urqmdparams_.u_at=AT1;  // Target identified
  urqmdparams_.u_zt=ZT1;
//------------------------------------------------------
//  identify Energy
//
      G4ThreeVector Pbefore = theTrack.Get4Momentum().vect();
      G4double T  = theTrack.GetKineticEnergy();
      G4double E  = theTrack.GetTotalEnergy();
      G4double TotalEbefore = E*AP1 +
      theTarget.AtomicMass(AT1, ZT1) + theTarget.GetEnergyDeposit();
//    ----------------------------------------------------------------

      if (AP1>1) {
      urqmdparams_.u_elab=T/(AP1*GeV); // Units are GeV/nuc for UrQMD

      E  = E/AP1;                // Units are GeV/nuc

      } else {

      urqmdparams_.u_elab=T/GeV;      //units are GeV

      TotalEbefore = E +
      theTarget.AtomicMass(AT1, ZT1) + theTarget.GetEnergyDeposit();
      }


//-------------------------------------------------------------
// identify impact parameter
  urqmdparams_.u_imp=-(1.1 * std::pow(G4double(AT1),(1./3.))); //units are
in fm for UrQMD;
//-------------------------------------------------------------
/////////////////////// initialise///////////////////

if (CurrentEvent==0)
{
G4cout << "\n creation of table, wait-------"<<G4endl;
```

```
G4cout << "\n"<<G4endl;

G4int io=0;

uinit_ (&io);


G4cout << "\n end to create  table "<<G4endl;

CurrentEvent=1;
}
////////////////////////////////////////////////////////

//#ifdef debug_G4UrQMD1_3Model

G4cout <<"UrQMDModel running-------------" <<G4endl;

urqmd_ ();

//#endif

//G4cout <<"Number of produced particles:  " <<sys_.npart<<G4endl;

G4int n              = sys_.npart;  //no of produced particles
if (n<2)
{
G4cout <<"===============Warning================"<<G4endl;
G4cout <<"======================================"<<G4endl;

G4cout <<"Number of produced particles is very low:  "
<<sys_.npart<<G4endl;
G4cout <<"============================================"<<G4endl;

//AND->
#if G4VERSION_NUMBER>=950
//New signature (9.5) for G4Exception
//Using G4HadronicException instead of base class
throw G4HadronicException(__FILE__,__LINE__,"Number of produced particle is
very low");
#else
G4Exception(" ");  //stop
#endif
//AND<-
} else {
  for (G4int i=0; i<n; i++)
  {


G4int pid=pdgid_ (&isys_.ityp[i], &isys_.iso3[i]);

// Particle is a final state secondary and not a nucleus.
// Determine what this secondary particle is, and if valid, load dynamic
// parameters.
//


  G4ParticleDefinition* pd=
  G4ParticleTable::GetParticleTable()->FindParticle(pid);
///////////////////////////////////////////////////////////////
//-----------------new_offshell
```

```
     if (offshell)
      {
////////--------------------------------   true off-shell!! --//
      if (pd)
      {


        G4double px        = (coor_.px[i]+ffermi_.ffermpx[i])* GeV;
//units are in MeV/c for G4
        G4double py        = (coor_.py[i]+ffermi_.ffermpy[i])* GeV;
        G4double pz        = (coor_.pz[i]+ffermi_.ffermpz[i])* GeV;

        G4double et        = (coor_.p0[i]) *GeV;


//    ------------------------------Use  "Lorentz vector"----------

     G4LorentzVector lorenzvec = G4LorentzVector(px,py,pz,et);

     cascadeParticle = new G4DynamicParticle(pd, lorenzvec);   //

     theResult.AddSecondary(cascadeParticle);
     }

      }else{

///////--better get rid of "off-mass-shell" nucleons---------//

      if (pd)
      {


      if (isys_.iso3[i]==1 || isys_.iso3[i]==-1)    //nucleons from UrQMD
      {
      if (coor_.fmass[i] >=0.938)    //units in GeV (UrQMD)
      {
        G4double px        = (coor_.px[i]+ffermi_.ffermpx[i])* GeV;
//units are in MeV/c for G4
        G4double py        = (coor_.py[i]+ffermi_.ffermpy[i])* GeV;
        G4double pz        = (coor_.pz[i]+ffermi_.ffermpz[i])* GeV;

        G4double et        = (coor_.p0[i]) *GeV;

//        G4double mass      = (coor_.fmass[i])*GeV;

//    ------------------------------Use  "Lorentz vector"----------

     G4LorentzVector lorenzvec = G4LorentzVector(px,py,pz,et);

     cascadeParticle = new G4DynamicParticle(pd, lorenzvec);   //

     theResult.AddSecondary(cascadeParticle);

      } //if mass

      } else {   //particles other than nucleons

        G4double px        = (coor_.px[i]+ffermi_.ffermpx[i])* GeV;
//units are in MeV/c for G4
        G4double py        = (coor_.py[i]+ffermi_.ffermpy[i])* GeV;
```

```
        G4double pz        = (coor_.pz[i]+ffermi_.ffermpz[i])* GeV;

        G4double et        = (coor_.p0[i]) *GeV;

//        G4double mass      = (coor_.fmass[i])*GeV;

//    ------------------------------Use "Lorentz vector"----------

     G4LorentzVector lorenzvec = G4LorentzVector(px,py,pz,et);

     cascadeParticle = new G4DynamicParticle(pd, lorenzvec);   //

     theResult.AddSecondary(cascadeParticle);


     } //else

//========================================================================
      }  //if pd


      } //else for off-shell



}  //for

} //if warning

//

//=========================================================================
if (verbose >= 3) {

//
    G4double TotalEafter = 0.0;
    G4ThreeVector TotalPafter;
    G4double charge    = 0.0;
    G4int baryon        = 0;
    G4int nSecondaries  = theResult.GetNumberOfSecondaries();

    for (G4int j=0; j<nSecondaries; j++) {
      TotalEafter += theResult.GetSecondary(j)->
        GetParticle()->GetTotalEnergy();

      TotalPafter += theResult.GetSecondary(j)->
        GetParticle()->GetMomentum();

      G4ParticleDefinition *pd = theResult.GetSecondary(j)->
        GetParticle()->GetDefinition();

      charge += pd->GetPDGCharge();
      baryon += pd->GetBaryonNumber();

    }  //for secondaries

    G4cout <<"----------------------------------------"
          <<"----------------------------------------"
          <<G4endl;
    G4cout <<"Total energy before collision  = " <<TotalEbefore    ///MeV
          <<" MeV" <<G4endl;
```

```
    G4cout <<"Total energy after collision    = " <<TotalEafter    //MeV
          <<" MeV" <<G4endl;

    G4cout <<"-----------------------------------------"<<G4endl;

    G4cout <<"Total momentum before collision = " <<Pbefore        //MeV
          <<" MeV/c" <<G4endl;
    G4cout <<"Total momentum after collision  = " <<TotalPafter    //MeV
          <<" MeV/c" <<G4endl;
    G4cout <<"-----------------------------------------"<<G4endl;

    if (verbose >= 4) {
      G4cout <<"Total charge before collision  = " <<(ZP+ZT)*eplus
            <<G4endl;
      G4cout <<"Total charge after collision    = " <<charge
            <<G4endl;

      G4cout <<"-----------------------------------------"<<G4endl;

      G4cout <<"Total baryon number before collision = "<<AP+AT
            <<G4endl;
      G4cout <<"Total baryon number after collision  = "<<baryon
            <<G4endl;
      G4cout <<"-----------------------------------------"<<G4endl;

    }  //if verbose4

    G4cout <<"-----------------------------------------"
          <<"-----------------------------------------"
          <<G4endl;

  }  //if verbose3


return &theResult;
}  //G4hadfinal


//-----------------------------------------------------------------------

//-----------------------------------------------------------------------
//
// WelcomeMessage
//
void G4UrQMD1_3Model::WelcomeMessage () const
{
  G4cout <<G4endl;
  G4cout <<"
******************************************************************"
        <<G4endl;
  G4cout <<" Interface to        G4UrQMD_1.3
activated"
        <<G4endl;
  G4cout <<" Version number : 00.00.0B          File date : 25/01/12"
<<G4endl;
  G4cout <<" (Interface written by Kh. Abdel-Waged et al. for the
KACST/NCMP)"
        <<G4endl;
  G4cout <<G4endl;
  G4cout <<"
******************************************************************"
```

```
        <<G4endl;
  G4cout << G4endl;

  return;
}

void G4UrQMD1_3Model::InitialiseDataTables ()
{
//
//
// The next line is to make sure the block data statements are
// executed.
//

g4urqmdblockdata_ ();


///////////////////////////////////////////////////
/////// Dynamic seed //////////////////////////////
//G4int ranseed=-time_ ();
//    Fixed seed  ///////////////////////////

G4int ranseed=1097569630;

G4cout <<"\n seed:  "<<ranseed<<G4endl;

sseed_ (&ranseed);

loginit_();

}
```

# 11 G4UrQMD1_3Model.hh file

```
//
// ********************************************************************
// * License and Disclaimer                                           *
// *                                                                  *
// * The  Geant4 software  is  copyright of the Copyright Holders  of *
// * the Geant4 Collaboration.  It is provided  under  the terms  and *
// * conditions of the Geant4 Software License,  included in the file *
// * LICENSE and available at  http://cern.ch/geant4/license .  These *
// * include a list of copyright holders.                             *
// *                                                                  *
// * Neither the authors of this software system, nor their employing *
// * institutes,nor the agencies providing financial support for this *
// * work  make  any representation or  warranty, express or implied, *
// * regarding  this  software system or assume any liability for its *
// * use.  Please see the license in the file  LICENSE  and URL above *
// * for the full disclaimer and the limitation of liability.         *
// *                                                                  *
// * This  code  implementation is the result of  the  scientific and *
// * technical work of the GEANT4 collaboration.                      *
// *                                                                  *
// * Parts of this code which have been  developed by Abdel-Waged     *
// * et al under contract (31-465) to the King Abdul-Aziz City for    *
// * Science and Technology (KACST), the National Centre of           *
// * Mathematics and Physics (NCMP), Saudi Arabia.                    *
// *                                                                  *
// * Rights to use, copy, modify and  redistribute this software for  *
// * general public use are granted in compliance with any licensing, *
// * distribution and development policy adopted by the Geant4        *
// * By using,  copying,  modifying or  distributing the software (or *
// * any work based  on the software)  you  agree  to acknowledge its *
// * use  in  resulting  scientific  publications,  and indicate your *
// * acceptance of all terms of the Geant4 Software license.          *
// ********************************************************************
//
#ifndef G4UrQMD1_3Model_hh
#define G4UrQMD1_3Model_hh
//
%%%%%%%%%%%%%%%%%%%%%%%%%%%%%%%%%%%%%%%%%%%%%%%%%%%%%%%%%%%%%%%%%%%%%%%%%%%%
%
//
// MODULE:            G4UrQMD1_3Model.hh
//
// Version:           0.B
// Date:          20/10/12
// Author:        Kh. Abdel-Waged and Nuha Felemban
// Revised by:        V.V. Uzhinskii
//                    SPONSERED BY
// Customer:            KACST/NCMP
// Contract:            31-465
//
//
%%%%%%%%%%%%%%%%%%%%%%%%%%%%%%%%%%%%%%%%%%%%%%%%%%%%%%%%%%%%%%%%%%%%%%%%%%%%
%
//
// Class Description
//
//
```

```
// Class Description - End
//
//
%%%%%%%%%%%%%%%%%%%%%%%%%%%%%%%%%%%%%%%%%%%%%%%%%%%%%%%%%%%%%%%%%%%%%%%%%%%%
%
////////////////////////////////////////////////////////////////////////////
////
#include "G4Nucleon.hh"
#include "G4Nucleus.hh"
#include "G4VIntraNuclearTransportModel.hh"
#include "G4KineticTrackVector.hh"
#include "G4FragmentVector.hh"
#include "G4ParticleChange.hh"
#include "G4ReactionProductVector.hh"
#include "G4ReactionProduct.hh"
#include "G4IntraNucleiCascader.hh"
#include "G4Track.hh"
#include <fstream>
#include <string>
//---------------



////////////////////////////////////////////////////////////////////////////
/////

class G4UrQMD1_3Model : public G4VIntraNuclearTransportModel {



public:

    G4UrQMD1_3Model(const G4String& name = "UrQMD1_3");


    virtual ~G4UrQMD1_3Model ();

//    G4double GetMinEnergy( const G4Material *aMaterial,
//                              const G4Element *anElement ) const;
//    G4double GetMaxEnergy( const G4Material *aMaterial,
//                              const G4Element *anElement ) const;


   G4ReactionProductVector* Propagate(G4KineticTrackVector*
   theSecondaries, G4V3DNucleus* theTarget);



    virtual G4HadFinalState *ApplyYourself
       (const G4HadProjectile &, G4Nucleus &);




   private:

   G4int operator==(G4UrQMD1_3Model& right) {
    return (this == &right);
   }

   G4int operator!=(G4UrQMD1_3Model& right) {
```

```
    return (this != &right);
   }


   G4int verbose;   //print options

   G4bool  offshell; //nucleons on- or off-shell

   void InitialiseDataTables();

  G4int CurrentEvent;

  private:


    void WelcomeMessage () const;

    G4HadFinalState theResult;


};

// inline G4double G4UrQMD1_3Model::GetMinEnergy( const G4Material *,
//   const G4Element * ) const
//   {return theMinEnergy;}
////////////////////////////////////////////////////////////////////////////
/////
//
// inline G4double G4UrQMD1_3Model::GetMaxEnergy( const G4Material *,
//   const G4Element * ) const
//   {return theMaxEnergy;}



////////////////////////////////////////////////////////////////////////////
/////
//
     #endif
```

# 12 G4UrQMD1_3Interface.hh file

```
//
// ********************************************************************
// * License and Disclaimer                                           *
// *                                                                  *
// * The  Geant4 software  is  copyright of the Copyright Holders  of *
// * the Geant4 Collaboration.  It is provided  under  the terms  and *
// * conditions of the Geant4 Software License,  included in the file *
// * LICENSE and available at  http://cern.ch/geant4/license .  These *
// * include a list of copyright holders.                             *
// *                                                                  *
// * Neither the authors of this software system, nor their employing *
// * institutes,nor the agencies providing financial support for this *
// * work  make  any representation or  warranty, express or implied, *
// * regarding  this  software system or assume any liability for its *
// * use.  Please see the license in the file  LICENSE  and URL above *
// * for the full disclaimer and the limitation of liability.         *
// *                                                                  *
// * This  code  implementation is the result of  the  scientific and *
// * technical work of the GEANT4 collaboration.                      *
// *                                                                  *
// * Parts of this code which have been  developed by Abdel-Waged     *
// * et al under contract (31-465) to the King Abdul-Aziz City for    *
// * Science and Technology (KACST), the National Centre of           *
// * Mathematics and Physics (NCMP), Saudi Arabia.
// *                                                                  *
// * By using,  copying,  modifying or  distributing the software (or *
// * any work based  on the software)  you  agree  to acknowledge its *
// * use  in  resulting  scientific  publications,  and indicate your *
// * acceptance of all terms of the Geant4 Software license.          *
// ********************************************************************
//
#ifndef G4UrQMD1_3Interface_hh
#define G4UrQMD1_3Interface_hh

//
%%%%%%%%%%%%%%%%%%%%%%%%%%%%%%%%%%%%%%%%%%%%%%%%%%%%%%%%%%%%%%%%%%%%%%%%%%%
%
//
// MODULE:           G4UrQMD1_3Model.hh
//
// Version:          0.B
// Date:         20/12/12
// Author:       Kh. Abdel-Waged and Nuha Felemban
// Revised by:       V.V. Uzhinskii
//                   SPONSERED BY
// Customer:           KACST/NCMP
// Contract:           31-465
//
//
%%%%%%%%%%%%%%%%%%%%%%%%%%%%%%%%%%%%%%%%%%%%%%%%%%%%%%%%%%%%%%%%%%%%%%%%%%%
%
//
//
// Class Description
//
//
// Class Description - End
//
```

```
//
%%%%%%%%%%%%%%%%%%%%%%%%%%%%%%%%%%%%%%%%%%%%%%%%%%%%%%%%%%%%%%%%%%%%%%%%%%%
%
///////////////////////////////////////////////////////////////////////////
////
#include "globals.hh"
//  coms
//
const G4int  nmax  = 500;
const G4int  nspl  = 500;
const G4int  smax  = 500;
//  comres
const G4int  minnuc=1;
const G4int  minmes=100;
const G4int  maxmes=132;
const G4int  numnuc=16;
const G4int  numdel=10;
const G4int  maxnuc=minnuc+numnuc-1;
const G4int  mindel=minnuc+maxnuc;
const G4int  maxdel=mindel+numdel-1;
const G4int  minres=minnuc+1;
const G4int  maxres=maxdel;
const G4int  numlam=13;
const G4int  numsig=9;
const G4int  numcas=6;
const G4int  numome=1;
const G4int  minlam=mindel+numdel;
const G4int  maxlam=minlam+numlam-1;
const G4int  minsig=minlam+numlam;
const G4int  maxsig=minsig+numsig-1;
const G4int  mincas=minsig+numsig;
const G4int  maxcas=mincas+numcas-1;
const G4int  minome=mincas+numcas;
const G4int  maxome=minome+numome-1;
const G4int  minbar=minnuc;
const G4int  maxbar=maxome;
const G4int  offmeson=minmes;
const G4int  maxmeson=maxmes;
const G4int  maxbra=11;
const G4int  maxbrm=25;
const G4int  maxbrs1=10;
const G4int  maxbrs2=3;
const G4int  nsigs = 10;
const G4int  itblsz= 100;
const G4int  maxreac = 13;
const G4int  maxpsig = 12;
//
//comwid
//
const G4int    widnsp=120;
const G4double mintab=0.10;
const G4double maxtab1=5.0;
const G4double maxtab2=50.0;
const G4int    tabver=9;
//
// options
//
const G4int numcto=400;
const G4int numctp=400;
const G4int maxstables=20;
//
```

```
// colltab (collision tables)
//
const G4int ncollmax = 100;
//
// inputs
//
const G4int aamax=300;
//
// newpart (new created particles)
//
const G4int  mprt=200;
const G4int  oprt=2;
//
// boxinc
//
const G4int bptmax=20;
//

// This next line is required as the default version of FORTRAN LOGICAL is
// four bytes long, whereas storage for G4bool is one byte.
//
// comnorm
const G4int n = 400;
//
// comstr
const G4int njspin=8;
//
//iso
const G4int jmax=7;

// This next line is required as the default version of FORTRAN LOGICAL is
// four bytes long, whereas storage for G4bool is one byte.
//

typedef G4int ftnlogical;

//
// Standard common block for UrQMD
// Common options for coms.f
//  20 commons
//
//
struct ccurqmd13urqmdparams
{
G4int u_at,u_zt,u_ap,u_zp;
G4double u_elab,u_imp;
G4int u_sptar,u_spproj;
};

struct ccurqmd13sys
{

G4int   npart, nbar, nmes, ctag,nsteps,uid_cnt,
        ranseed,event,ap,at,zp,zt,eos,dectag,
        nhardres, nsoftres, ndecres, nelcoll, nblcoll;
};


struct ccurqmd13rsys
{
G4double time,acttime,bdist,bimp,bmin,ebeam,ecm;
```

```
};

struct ccurqmd13comseed
{
  ftnlogical
            firstseed;
};



struct ccurqmd13logic
{
  ftnlogical
            lsct[nmax], logSky, logYuk, logCb, logPau;

};

struct ccurqmd13mdprop
{
 G4double
        r0_t[nmax], rx_t[nmax], ry_t[nmax], rz_t[nmax];
};


struct ccurqmd13cuts
{
  G4double
          cutmax, cutPau, cutCb, cutYuk, cutSky, cutdww;
};


struct ccurqmd13spdata
{
  G4double
          spx[nspl], spPauy[nspl], outPau[nspl],
          spCby[nspl],  outCb[nspl],
          spYuky[nspl], outYuk[nspl],
          spSkyy[nspl], outSky[nspl],
          spdwwy[nspl], outdww[nspl];
};


struct ccurqmd13isys
{

G4int spin[nmax],ncoll[nmax],charge[nmax],ityp[nmax],
      lstcoll[nmax],
      iso3[nmax],origin[nmax],strid[nmax],uid[nmax];
};


struct ccurqmd13coor
{
G4double
        r0[nmax], rx[nmax], ry[nmax], rz[nmax],
        p0[nmax], px[nmax], py[nmax], pz[nmax],
        fmass[nmax], rww[nmax],dectime[nmax];
};

struct ccurqmd13frag
{
```

```
G4double
        tform[nmax], xtotfac[nmax];
};

struct ccurqmd13aios
{
G4double
        airx[nmax], airy[nmax], airz[nmax],
        aipx[nmax], aipy[nmax], aipz[nmax],
        aorx [4][nmax], aory[4][nmax], aorz[4][nmax],
        aopx[4][nmax], aopy[4][nmax], aopz[4][nmax];
};


struct ccurqmd13pots
{

G4double
       Cb0, Yuk0, Pau0, Sky20, Sky30, gamSky,
       gamYuk, drPau, dpPau, gw, sgw, delr, fdel,
       dt,da, db,dtimestep;
};

struct ccurqmd13scoor
{

G4double
     r0s[smax], rxs[smax], rys[smax], rzs[smax],
     p0s[smax], pxs[smax], pys[smax], pzs[smax],
     sfmass[smax];
};

struct ccurqmd13sisys
{
  G4int
      sspin[smax], scharge[smax], sityp[smax], siso3[smax],
     suid[smax];
};


struct ccurqmd13ssys
{
  G4int  nspec;
};

struct ccurqmd13rtdelay
{
G4double
        p0td[nmax][2],pxtd[nmax][2],pytd[nmax][2],pztd[nmax][2],
        fmasstd[nmax][2];
};


struct ccurqmd13itdelay
{
G4int
    ityptd[nmax][2],iso3td[nmax][2];
};

struct ccurqmd13svinfo
{
```

```
G4int
     itypt[2],uidt[2],origint[2],iso3t[2];
};


struct ccurqmd13ffermi
{
G4double
       ffermpx[nmax], ffermpy[nmax], ffermpz[nmax];
};

struct ccurqmd13peq
{
G4double peq1, peq2;
};

//
// Definition for Collision Term
// Commons  comres
// 4 commons
//

struct ccurqmd13versioning
{
char versiontag[45];
};


struct ccurqmd13resonances
{

      G4double massres[maxbar-minbar+1],widres[maxbar-minbar+1];
      G4double massmes[maxmes-minmes+1];
      G4double widmes[maxmes-minmes+1];
      G4double mmesmn[maxmes-minmes+1];
      G4double branres[maxdel-minnuc][maxbra+1];
      G4double branmes[maxmes-minmes][maxbrm+1];

      G4double branbs1[maxsig-minlam][maxbrs1+1];
      G4double branbs2[maxcas-mincas][maxbrs2+1];

      G4int  bs1type[maxbrs1+1][4],bs2type[maxbrs2+1][4];
      G4int lbs1[maxsig-minlam][maxbrs1+1];
      G4int lbs2[maxcas-mincas][maxbrs2+1];
      G4int lbm[maxmes-minmes][maxbrm+1];

      G4int  jres[maxbar-minbar+1];
      G4int  jmes[maxmes-minmes+1];
      G4int lbr[maxdel-minnuc][maxbra+1];
      G4int  brtype[maxbra+1][4];
      G4int  pares[maxbar-minbar+1],pames[maxmes-minmes+1];
      G4int  bmtype[maxbrm+1][4];
      G4int  isores[maxbar-minbar+1], isomes[maxmes-minmes+1];
      G4int  strres[maxbar-minbar+1],strmes[maxmes-minmes+1];
      G4int mlt2it[maxmes-minmes];
};

struct ccurqmd13sigtabi
{
G4int sigmaln[maxreac][2][maxpsig];
G4int sigmainf[20][nsigs];
```

```
};


struct  ccurqmd13sigtabr
{

G4double  sigmas[itblsz][nsigs],sigmascal[5][nsigs];
};

//comwid
struct ccurqmd13decaywidth
{
G4double tabx [widnsp];
G4double fbtaby [2][maxbar-minbar+1][widnsp];
G4double  pbtaby[maxbra+1][maxbar-minbar+1][2][widnsp];
G4double  fmtaby [2][maxmes-minmes+1][widnsp];
G4double  pmtaby [maxbrm+1][maxmes-minmes+1][2][widnsp];
G4int     wtabflg;

};


struct ccurqmd13brwignorm
{
G4double bwbarnorm[maxbar-minbar+1];
G4double bwmesnorm[maxmes-minmes+1];
};


struct ccurqmd13xsections
{
G4double tabxnd [widnsp];
G4double frrtaby[maxdel-1][2][2][widnsp];
};


struct ccurqmd13tabnames
{
char tabname[77];
};
//------------------
//
// options
//
struct ccurqmd13options
{
G4int    CTOption[numcto];

G4double CTParam[numctp];
};


struct ccurqmd13optstrings
{
char ctodc[numcto][2];
char ctpdc[numctp][2];
};

struct ccurqmd13loptions
{
ftnlogical
```

```
         fixedseed,bf13,bf14,bf15,bf16,bf17,bf18,bf19,
         bf20;
};

struct ccurqmd13stables
{
G4int nstable;
G4int stabvec[maxstables];
};

//
//colltab
//
struct ccurqmd13colltab
{
G4double
           cttime[ncollmax+1],ctsqrts[ncollmax],
           ctsigtot[ncollmax],tmin;
G4int
           cti1[ncollmax],cti2[ncollmax];
G4int
           nct,actcol;
ftnlogical
           ctvalid[ncollmax];
G4int
           ctsav[ncollmax];
G4int
           nsav,apt;
G4double
           ctcolfluc[ncollmax];
};

//
// inputs
//
struct ccurqmd13inputs
{
G4int  nevents,spityp[2],prspflg;
G4int  trspflg,spiso3[2],outsteps,bflag,srtflag,efuncflag;
G4int  nsrt,firstev,npb;
};

struct ccurqmd13input2
{
G4double srtmin,srtmax,pbeam,betann,betatar,betapro;
G4double pbmin,pbmax;
};

struct ccurqmd13protarints
{

G4int pt_iso3[2][aamax],pt_ityp[2][aamax],pt_spin[2][aamax];

G4int pt_charge[2][aamax],pt_aa[2],pt_uid[2][aamax];
};

struct ccurqmd13protarreals
{
G4double pt_r0[2][aamax],pt_rx[2][aamax],pt_ry[2][aamax],
         pt_rz[2][aamax],pt_fmass[2][aamax],pt_dectime[2][aamax];
G4double pt_p0[2][aamax],pt_px[2][aamax],pt_py[2][aamax],
```

```
         pt_pz[2][aamax];
G4double pt_rho[2][aamax];
G4double pt_pmax[2][aamax];
};
// newpart
struct ccurqmd13inewpart
{
G4int itypnew[mprt],i3new[mprt],itot[mprt],inew[mprt],nexit;
G4int iline,strcount,pslot[oprt],nstring1, nstring2,
      sidnew[mprt],itypold[oprt],iso3old[oprt];
};

struct ccurqmd13rnewpart
{
G4double  pnew[mprt][5],xnew[mprt][4],betax,betay,betaz,
          pold[oprt][5],p0nn,pxnn,pynn,pznn,pnn, mstring[2],
          pnnout,xtotfacold[oprt];

};

struct ccurqmd13fnewpart
{
G4double leadfac[mprt];
};
//
// boxinc
//
struct ccurqmd13boxic
{
G4int cbox;
G4int boxflag;
G4int mbox;
G4int bptityp[bptmax],bptiso3[bptmax],bptpart[bptmax];
G4int edensflag,para,solid, mbflag,mtest;
};

struct ccurqmd13boxrc
{
G4double bptpmax[bptmax];
G4double edens;
G4double lbox;
G4double lboxhalbe;
G4double lboxd;
G4double mbp0, mbpx, mbpy, mbpz;
};
// comnorm
struct ccurqmd13normsplin
{
G4double x_norm[n][4],y_norm[n][4];
G4double y2a[n][4],y2b[n][4], dx;
};
// comstr
struct ccurqmd13FRGSPA
{
G4double
          pjspns, pmix1s[njspin][3], pmix2s[njspin][3]
          , pbars, parqls, parrs;
};
struct ccurqmd13FRGCPA
{
G4double
```

```
          pjspnc, pmix1c[njspin][3], pmix2c[njspin][3], pbarc;
};

struct ccurqmd13coparm
{
G4double parm[njspin];
};

struct ccurqmd13const
{
G4double pi;
};
//// freezeout
//
struct ccurqmd13frcoor
{
G4double frr0[nmax], frrx[nmax], frry[nmax], frrz[nmax],
         frp0[nmax], frpx[nmax], frpy[nmax], frpz[nmax];
};

//  input
struct ccurqmd13values
{
G4double valint[1];
};

// cascinit
struct ccurqmd13ini
{
ftnlogical bcorr;
};


// iso
struct ccurqmd13factorials
{
G4double logfak[101];
};
struct ccurqmd13cgks
{

G4double cgktab[jmax+1][2*jmax+1][2*jmax+1][jmax+1][jmax+1];
};

// UrQMD
//
struct ccurqmd13energies
{
G4double ekinbar, ekinmes, esky2, esky3,
         eyuk, ecb, epau;
};



// urqmd
extern "C"
{
extern int time_ ();
extern void loginit_();
extern void sseed_ (int*);
extern void uinit_ (int*);
```

```
extern void urqmd_ ();
extern int pdgid_ (int*, int*); //ityp

extern void g4urqmdblockdata_ ();

// urqmdparams
extern struct ccurqmd13urqmdparams  urqmdparams_;
//coms
extern struct ccurqmd13sys     sys_;
extern struct ccurqmd13rsys    rsys_;
extern struct ccurqmd13comseed comseed_;
extern struct ccurqmd13logic   logic_;
extern struct ccurqmd13mdprop  mdprop_;
extern struct ccurqmd13cuts    cuts_;
extern struct ccurqmd13spdata  spdata_;
extern struct ccurqmd13isys    isys_;
extern struct ccurqmd13coor    coor_;
extern struct ccurqmd13frag    frag_;
extern struct ccurqmd13aios    aios_;
extern struct ccurqmd13pots    pots_;
extern struct ccurqmd13scoor   scoor_;
extern struct ccurqmd13sisys   sisys_;
extern struct ccurqmd13ssys    ssys_;
extern struct ccurqmd13rtdelay rtdelay_;
extern struct ccurqmd13itdelay itdelay_;
extern struct ccurqmd13svinfo  svinfo_;
extern struct ccurqmd13ffermi  ffermi_;
extern struct ccurqmd13peq     peq_;
//comres
extern struct ccurqmd13versioning  versioning_;
extern struct ccurqmd13resonances  resonances_;
extern struct ccurqmd13sigtabi  sigtabi_;
extern struct ccurqmd13sigtabr sigtabr_;

//comwid
extern struct ccurqmd13decaywidth decaywidth_;
extern struct ccurqmd13brwignorm  brwignorm_;
extern struct ccurqmd13xsections  xsections_;
extern struct ccurqmd13tabnames   tabnames_;
//options
extern struct ccurqmd13options    options_;
extern struct ccurqmd13optstrings optstrings_;
extern struct ccurqmd13loptions    loptions_;
extern struct ccurqmd13stables     stables_;
//colltab
extern struct ccurqmd13colltab     colltab_;
//inputs
extern struct ccurqmd13inputs      inputs_;
extern struct ccurqmd13input2      input2_;
extern struct ccurqmd13protarints  protarints_;
extern struct ccurqmd13protarreals protarreals_;
//newpart
extern struct ccurqmd13inewpart    inewpart_;
extern struct ccurqmd13rnewpart    rnewpart_;
extern struct ccurqmd13fnewpart    fnewpart_;
//bocinc
extern struct ccurqmd13boxic       boxic_;
extern struct ccurqmd13boxrc       boxrc_;
// comnorm
struct ccurqmd13normsplin  normsplin_;
//comstr
```

```
struct ccurqmd13FRGSPA  FRGSPA_;
struct ccurqmd13FRGCPA  FRGCPA_;
struct ccurqmd13coparm  coparm_;
struct ccurqmd13const   const_;
// freezeout
struct ccurqmd13frcoor  frcoor_;
//urqmd
extern struct ccurqmd13energies  energies_;
//input
extern struct ccurqmd13values values_;
// cascinit
extern struct ccurqmd13ini ini_;
//iso
extern struct ccurqmd13factorials factorials_;
extern struct ccurqmd13cgks  cgks_;

}

#endif
```

## Acknowledments

Kh. Abdel-Waged and N. Felemban would like to thank geant4 group, especially Dr Andrea Dotti for his efforts in making the software compatible with geant4-9.5. This work is supported by the King Abdul-Aziz City for Science and Technology, the National Centre of Mathematics and Physics, Saudi Arabia, contract number 31-465. Kh Abdel-Waged and N. Felemban are thankful to the members of Geant4 hadronic group for stimulating discussions and help.